\documentclass[12pt]{article}
\usepackage{amsmath,amssymb}
\usepackage{bm}
\usepackage{mathrsfs}
\usepackage{fourier}
\usepackage{multirow}
\allowdisplaybreaks[4]
\newcommand{{\SlashD}}{D\!\!\!\!\!\!\big/}
\newcommand{{\Slashq}}{q\!\!\!\!\!\big/}
\newcommand{{\SlashF}}{{\rm F}\!\!\!\!\!/}
\usepackage[usenames]{color}

\begin{document}

\title{Flavor structure
from `canonical' Yukawa interactions
and `emergent' kinetic terms}

\author{
Yoshiharu \textsc{Kawamura}\footnote{E-mail: haru@azusa.shinshu-u.ac.jp}\\
{\it Department of Physics, Shinshu University, }\\
{\it Matsumoto 390-8621, Japan}\\
}


\maketitle
\begin{abstract}
We study the flavor structure of quarks in the standard model
from a viewpoint of a canonical type of Yukawa interactions 
and an emergence of kinetic terms.
A realistic structure can be generated based on the emergence proposal 
that quark kinetic terms appear in the infra-red region,
as a result of radiative corrections
involving towers of massive states.
\end{abstract}

\section{Introduction}

The origin of the fermion mass hierarchy and flavor mixing in the standard model (SM)
has been a big enigma.
The reason why it is difficult to uncover the flavor structure
is that Yukawa interactions in the SM contain many unobservable parameters,
which are eliminated by bi-unitary transformations
of global symmetries on fermion kinetic terms, 
and useful information to determine a physics beyond the SM
is not fully obtained from precision measurements of the SM parameters alone. 

If the coexistence of matter kinetic terms and Yukawa interactions in the SM
and the appearance of flavor symmetries on the kinetic terms
complicate an understanding of the flavor structure,
it must be better to return to the origin of each term in the SM Lagrangian density.
Hence, we go with the idea that the origin of kinetic terms and Yukawa interactions
can give a key to solve the enigma.

In the usual case, we assume that kinetic terms (including gauge fields
via gauge interactions) exist from the beginning,
and then chiral fermion fields are determined, 
up to some global unitary transformation,
by making their kinetic terms the canonical ones.
We cast doubt on it by considering a case that kinetic terms are absent
and Yukawa interactions such as
$y_{ij} \overline{\chi}_{{\rm L}i} \varphi \eta_{{\rm R}j}$ are present
in the ultra-violet (UV) region at a fundamental theory level.
Here, $y_{ij}$ is a Yukawa coupling matrix, $i, j(=1, 2, 3)$ are family labels,
summation over repeated indices is understood, in most cases, throughout this paper,
${\chi}_{{\rm L}i}$ and $\eta_{{\rm R}j}$ are chiral fermions 
and $\varphi$ is a scalar field.
In such a case, chiral fermion fields can be defined through Yukawa interactions
and the simplest choice of the Yukawa coupling matrix is $y_{ij} = \delta_{ij}$
($\delta_{ij}$ equals to 1 for $i=j$ and 0 for $i \ne j$).
We refer to this type of Yukawa interactions as `canonical' Yukawa interactions.
We note that there is a freedom to change $y_{ij}$ into $\delta_{ij}$ 
if $y_{ij}$ has non-zero singular values,
although fields are not uniquely fixed 
and couplings to extra fields can become complicated.
Then, the problem on the flavor structure is transported into
that on the origin of kinetic terms, under the assumption that 
kinetic terms appear in the infra-red (IR) region at an effective theory level.
We refer to this kind of kinetic terms as `emergent' kinetic terms.
In this way, the flavor structure in the SM can be originated from
a counterpart in the emergent kinetic terms.

Recently, a generation of kinetic terms with large coefficients
has been proposed based on the emergence proposal in the strong version
that {\it ``In a theory of quantum gravity, all light fields in a perturbative regime
have no kinetic terms in the UV. The required kinetic terms appear as an IR effect
after integrating out towers of massive states below the quantum gravity
cut-off scale.''}~\cite{H,GPV,HRR,HRR2,P},
and its phenomenological implications including the fermion mass hierarchy
and the electro-weak hierarchy problem have been studied~\cite{CHI,CHI2}.
The emergence proposal has been presented as part of the Swampland program~\cite{V}.

In this paper, we study the flavor structure of quarks in the SM
from a viewpoint of canonical Yukawa interactions and emergent kinetic terms
and examine whether a realistic structure can be generated or not
based on the above emergence proposal. 

The outline of this paper is as follows.
In the next section, we review the flavor structure of quarks
and study a structure of kinetic terms based on canonical Yukawa interactions in the SM.
In Sect. 3, we investigate a generation of quark kinetic terms
and a formation of the flavor structure, using a simple model.
In the last section, we give conclusions and discussions.

\section{`Canonical' Yukawa interactions}

First, we review the quark sector in the SM, based on the usual Lagrangian density:
\begin{eqnarray}
&~& \mathscr{L}_{\rm SM}^{\rm quark} = \overline{q}_{{\rm L}i} i \SlashD q_{{\rm L}i}
+ \overline{u}_{{\rm R}i} i \SlashD u_{{\rm R}i}
+ \overline{d}_{{\rm R}i} i \SlashD d_{{\rm R}i}
\nonumber \\
&~& ~~~~~~~~~~~~~~~~~
- y_{ij}^{(u)} \overline{q}_{{\rm L}i} \tilde{\phi} u_{{\rm R}j}
- y_{ij}^{(d)} \overline{q}_{{\rm L}i} \phi d_{{\rm R}j} + {\rm h.c.},
\label{Lq-SM}
\end{eqnarray}
where $q_{{\rm L}i}$ are left-handed quark doublets,
$u_{{\rm R}i}$ and $d_{{\rm R}i}$ are right-handed up- and down-type quark singlets,
$i, j=1,2,3$,
$y_{ij}^{(u)}$ and $y_{ij}^{(d)}$ are Yukawa coupling matrices,
$\phi$ is the Higgs doublet, $\tilde{\phi} = i \tau_2 \phi^*$
and h.c. stands for hermitian conjugation of former terms.
The Yukawa coupling matrices are diagonalized 
as $V_{\rm L}^{(u)} y^{(u)} {V_{\rm R}^{(u)}}^{\dagger} = y_{\rm diag}^{(u)}$
and $V_{\rm L}^{(d)} y^{(d)} {V_{\rm R}^{(d)}}^{\dagger} = y_{\rm diag}^{(d)}$
by bi-unitary transformations 
and the quark masses are obtained as
\begin{eqnarray}
&~& V_{\rm L}^{(u)} y^{(u)} {V_{\rm R}^{(u)}}^{\dagger} \frac{v}{\sqrt{2}}
= y_{\rm diag}^{(u)} \frac{v}{\sqrt{2}} = {\rm diag}\left(m_u, m_c, m_t\right),
\label{Mu-diag}\\
&~& V_{\rm L}^{(d)} y^{(d)} {V_{\rm R}^{(d)}}^{\dagger} \frac{v}{\sqrt{2}}
= y_{\rm diag}^{(d)} \frac{v}{\sqrt{2}} = {\rm diag}\left(m_d, m_s, m_b\right),
\label{Md-diag}
\end{eqnarray}
where $V_{\rm L}^{(u)}$, $V_{\rm L}^{(d)}$, $V_{\rm R}^{(u)}$ and $V_{\rm R}^{(d)}$
are unitary matrices, $v/\sqrt{2}$ is the vacuum expectation value of neutral component 
in the Higgs doublet, family labels are omitted,
and $m_u$, $m_c$, $m_t$, $m_d$, $m_s$ and $m_b$ are masses of up, charm, top,
down, strange and bottom quarks, respectively.

As seen from eqs.~\eqref{Mu-diag} and \eqref{Md-diag}, 
the quark Yukawa coupling matrices are expressed by
\begin{eqnarray}
y^{(u)} = {V_{\rm L}^{(u)}}^{\dagger} y_{\rm diag}^{(u)} V_{\rm R}^{(u)},~~
y^{(d)} = {V_{\rm L}^{(d)}}^{\dagger} y_{\rm diag}^{(d)} V_{\rm R}^{(d)}
= {V_{\rm L}^{(u)}}^{\dagger} V_{\rm CKM} y_{\rm diag}^{(d)} V_{\rm R}^{(d)},
\label{y-quark}
\end{eqnarray}
using $V_{\rm L}^{(u)}$, $V_{\rm R}^{(u)}$, $V_{\rm R}^{(d)}$,
$y_{\rm diag}^{(u)}$, $y_{\rm diag}^{(d)}$
and the Cabibbo-Kobayashi-Maskawa matrix defined by~\cite{C,KM}
\begin{eqnarray}
V_{\rm CKM} \equiv V_{\rm L}^{(u)} {V_{\rm L}^{(d)}}^{\dagger}.
\label{VKM}
\end{eqnarray}
The $3 \times 3$ matrices $V_{\rm L}^{(u)}$,
$V_{\rm R}^{(u)}$ and $V_{\rm R}^{(d)}$ are completely unknown in the SM,
because they can be eliminated by the global 
${\rm U}(3) \times {\rm U}(3) \times {\rm U}(3)/{\rm U}(1)$ symmetry
that the quark kinetic terms possess.
Here, a global $U(1)$ phase is constrained by a global $U(1)$ invariance
in Yukawa interactions.
We have a situation that information on a physics beyond the SM 
is not fully obtained by observable parameters 
such as $y_{\rm diag}^{(u)}$, $y_{\rm diag}^{(d)}$ and $V_{\rm CKM}$ alone.

In Table \ref{Ty}, the number of independent parameters 
relating to $y^{(u)}_{ij}$ and $y^{(d)}_{ij}$ is listed.
\begin{table}[htb]
\begin{center}
\begin{tabular}{|c|c|} \hline
Yukawa couplings and related ones & Number of parameters \\ \hline\hline
$y^{(u)}_{ij}$,~~ $y^{(d)}_{ij}$ & $18$,~~ $18$ \\ \hline
$y_{\rm diag}^{(u)}$,~~ $y_{\rm diag}^{(d)}$ & $3$,~~ $3$ \\
$V_{\rm CKM}$ & $4 (=3+1)$ \\
$V_{\rm L}^{(u)}$,~~ $V_{\rm R}^{(u)}$,~~ $V_{\rm R}^{(d)}$ 
& $26 (=9+9+9-1)$ \\ \hline
\end{tabular}
\end{center}
\caption[table-y]{The number of parameters relating to quark Yukawa couplings}
\label{Ty}
\end{table}
As $y^{(u)}_{ij}$ and $y^{(d)}_{ij}$ are $3 \times 3$ complex matrices, 
they totally have 36 parameters.
The $V_{\rm CKM}$ contains three mixing angles and a CP violating phase.
The $V_{\rm L}^{(u)}$, $V_{\rm R}^{(u)}$ and $V_{\rm R}^{(d)}$ own
26 unobservable parameters on the global 
${\rm U}(3) \times {\rm U}(3) \times {\rm U}(3)/{\rm U}(1)$ symmetry.

Next, we consider an unusual case with `canonical' Yukawa interactions given by
\begin{eqnarray}
\mathscr{L}_{\rm CY}^{\rm quark} 
=  - \overline{q}'_{{\rm L}i} \tilde{\phi} u'_{{\rm R}i}
- \overline{q}'_{{\rm L}i} \phi d'_{{\rm R}i} + {\rm h.c.},
\label{Lq-CY}
\end{eqnarray}
in preparation for the study on the emergence of kinetic terms in the next section.
Using the field variables with a prime, 
the Lagrangian density in the quark sector is written by
\begin{eqnarray}
&~& {\mathscr{L}'}_{\rm SM}^{\rm quark} 
= k_{ij}^{(q)} \overline{q}'_{{\rm L}i} i \SlashD q'_{{\rm L}j}
+ k_{ij}^{(u)} \overline{u}'_{{\rm R}i} i \SlashD u'_{{\rm R}j}
+ k_{ij}^{(d)} \overline{d}'_{{\rm R}i} i \SlashD d'_{{\rm R}j}
\nonumber \\
&~& ~~~~~~~~~~~~~~~~~~~~ - \overline{q}'_{{\rm L}i} \tilde{\phi} u'_{{\rm R}i}
- \overline{q}'_{{\rm L}i} \phi d'_{{\rm R}i} + {\rm h.c.},
\label{L'q-SM}
\end{eqnarray}
where $k_{ij}^{(q)}$, $k_{ij}^{(u)}$ and $k_{ij}^{(d)}$ are 
kinetic coefficient matrices denoted by
\begin{eqnarray}
&~& k_{ij}^{(q)} = \left(\left(W^{(u)}\right)^{-1}
\left({W^{(u)}}^{\dagger}\right)^{-1}\right)_{ij},
\label{kq}\\
&~& k_{ij}^{(u)} = \left({W^{(u)}}^{\dagger} 
\left(y_{\rm diag}^{(u) -1}\right)^2 W^{(u)}\right)_{ij},
\label{ku}\\
&~& k_{ij}^{(d)} = \left({W^{(u)}}^{\dagger} 
V_{\rm CKM} \left(y_{\rm diag}^{(d) -1}\right)^2 
V_{\rm CKM}^{\dagger} W^{(u)}\right)_{ij}.
\label{kd}
\end{eqnarray}
Here, $W^{(u)}$ is a $3 \times 3$ complex matrix.
Note that non-canonical quark kinetic terms 
appear in $ {\mathscr{L}'}_{\rm SM}^{\rm quark}$.\footnote{
Several works on the flavor physics have been carried out
based on matter kinetic 
terms~\cite{GNM,I,BD1,BD2,BLR,NS,KY,HKY,KP,EI,KPRVV,Liu,DIU,PS,DGPP,YK,YK2,YK3}.}
Because $W^{(u)}$ is an arbitrary matrix, 
the expressions \eqref{kq} -- \eqref{kd} are not unique.
For instance, we obtain the relations:
\begin{eqnarray}
&~& k_{ij}^{(q)} = \left(\left(\widetilde{W}^{(u)}\right)^{-1}
\left(y_{\rm diag}^{(u) -1}\right)^2
\left(\widetilde{W}^{(u)\dagger}\right)^{-1}\right)_{ij},
\label{kq-tilde}\\
&~& k_{ij}^{(u)} = \left(\widetilde{W}^{(u)\dagger} 
\widetilde{W}^{(u)}\right)_{ij},
\label{ku-tilde}\\
&~& k_{ij}^{(d)} = \left(\widetilde{W}^{(u)\dagger}
y_{\rm diag}^{(u)} 
V_{\rm CKM} \left(y_{\rm diag}^{(d) -1}\right)^2 
V_{\rm CKM}^{\dagger} y_{\rm diag}^{(u)} \widetilde{W}^{(u)}\right)_{ij},
\label{kd-tilde}
\end{eqnarray}
using a $3 \times 3$ complex matrix $\widetilde{W}^{(u)}=y_{\rm diag}^{(u) -1}W^{(u)}$.
As another choice, we have the relations:
\begin{eqnarray}
&~& k_{ij}^{(q)} = \left(\left(W^{(d)}\right)^{-1}
\left({W^{(d)}}^{\dagger}\right)^{-1}\right)_{ij},
\label{kq(d)}\\
&~& k_{ij}^{(u)} = \left({W^{(d)}}^{\dagger}
V_{\rm CKM}^{\dagger} \left(y_{\rm diag}^{(u) -1}\right)^2 V_{\rm CKM} W^{(d)}\right)_{ij},
\label{ku(d)}\\
&~& k_{ij}^{(d)} = \left({W^{(d)}}^{\dagger} 
\left(y_{\rm diag}^{(d) -1}\right)^2 W^{(d)}\right)_{ij},
\label{kd(d)}
\end{eqnarray}
where $W^{(d)}=V_{\rm CKM}^{\dagger}W^{(u)}$.
We find that a seed of the mass hierarchy and flavor mixing can be hidden 
in various places.

In Table \ref{Tk}, the number of independent parameters 
concerning $k_{ij}^{(q)}$, $k_{ij}^{(u)}$ and $k_{ij}^{(d)}$ is listed.
\begin{table}[htb]
\begin{center}
\begin{tabular}{|c|c|} \hline
Kinetic coefficients and related ones & Number of parameters \\ \hline\hline
$k_{ij}^{(q)}$,~~ $k_{ij}^{(u)}$,~~ $k_{ij}^{(d)}$ & $9$,~~ $9$,~~ $9$ \\ \hline
$y_{\rm diag}^{(u)}$,~~ $y_{\rm diag}^{(d)}$ & $3$,~~ $3$ \\
$V_{\rm CKM}$ & $4 (=3+1)$ \\
$W^{(u)}$ & $17 (=18-1)$ \\ \hline
\end{tabular}
\end{center}
\caption[table-k]{The number of parameters relating to quark kinetic coefficients}
\label{Tk}
\end{table}
The $k_{ij}^{(q)}$, $k_{ij}^{(u)}$ and $k_{ij}^{(d)}$ totally have 27 
independent parameters because they are hermitian matrices.
Note that a global $U(1)$ phase in $W^{(u)}$
is canceled out and does not appear in eqs.~\eqref{kq} -- \eqref{kd}
and then the total number of independent parameters in $W^{(u)}$ is 17.

Let us show that ${\mathscr{L}'}_{\rm SM}^{\rm quark}$ 
is equivalent to $\mathscr{L}_{\rm SM}^{\rm quark}$.
The $k_{ij}^{(q)}$, $k_{ij}^{(u)}$ and $k_{ij}^{(d)}$ are also written by
\begin{eqnarray}
k^{(q)}_{ij} = \left(X_{q}^{\dagger}X_{q}\right)_{ij},~~
k^{(u)}_{ij} = \left(X_{u}^{\dagger}X_{u}\right)_{ij},~~
k^{(d)}_{ij} = \left(X_{d}^{\dagger}X_{d}\right)_{ij},
\label{kX}
\end{eqnarray}
where $3 \times 3$ complex matrices $X_q$, $X_u$ and $X_d$ parametrized by
\begin{eqnarray}
&~& X_q = {V_{\rm L}^{(u)}}^{\dagger} \left({W^{(u)}}^{\dagger}\right)^{-1},~~ 
\label{Xq}\\
&~& X_u = {V_{\rm R}^{(u)}}^{\dagger} \left(y^{(u)}_{\rm diag}\right)^{-1} W^{(u)},~~
\label{Xu}\\
&~& X_d = {V_{\rm R}^{(d)}}^{\dagger} \left(y^{(d)}_{\rm diag}\right)^{-1} 
V_{\rm CKM}^{\dagger} W^{(u)},
\label{Xd}
\end{eqnarray}
using eqs.~\eqref{kq} -- \eqref{kd}.
Note that unitary matrices $V_{\rm L}^{(u)}$, $V_{\rm R}^{(u)}$ and $V_{\rm R}^{(d)}$
made of unobservable parameters appear.
Using $X_q$, $X_u$ and $X_d$, the quarks $q_{\rm L}$, $u_{\rm R}$ and $d_{\rm R}$ 
in ${\mathscr{L}}_{\rm SM}^{\rm quark}$ are related to 
$q'_{\rm L}$, $u'_{\rm R}$ and $d'_{\rm R}$
in ${\mathscr{L}'}_{\rm SM}^{\rm quark}$ such that
\begin{eqnarray}
q_{\rm L} = X_q  q'_{\rm L},~~u_{\rm R} = X_u u'_{\rm R},~~d_{\rm R} = X_d d'_{\rm R}.
\label{qq'}
\end{eqnarray}
Using eqs.~\eqref{Xq} -- \eqref{qq'} and \eqref{y-quark}, 
the canonical Yukawa interactions are rewritten as
\begin{eqnarray}
\hspace{-1.2cm}
&~& \mathscr{L}_{\rm CY}^{\rm quark} 
=  - \overline{q}'_{{\rm L}i} \tilde{\phi} u'_{{\rm R}i}
- \overline{q}'_{{\rm L}i} \phi d'_{{\rm R}i} + {\rm h.c.} 
\nonumber \\
\hspace{-1.2cm}
&~& ~~~~~~~~~~~~~~~\! 
= - \overline{q}_{{\rm L}i} \left(\left(X_q^{\dagger}\right)^{-1} X_u^{-1}\right)_{ij}
\tilde{\phi}u_{{\rm R}j}
- \overline{q}_{{\rm L}i} \left(\left(X_q^{\dagger}\right)^{-1} X_d^{-1}\right)_{ij}
\phi d_{{\rm R}j} + {\rm h.c.} 
\nonumber \\
\hspace{-1.2cm}
&~& ~~~~~~~~~~~~~~~\! 
= - \overline{q}_{{\rm L}i} 
\left({V_{\rm L}^{(u)}}^{\dagger} y_{\rm diag}^{(u)} V_{\rm R}^{(u)}\right)_{ij}
\tilde{\phi}u_{{\rm R}j}
- \overline{q}_{{\rm L}i} 
\left({V_{\rm L}^{(u)}}^{\dagger} V_{\rm CKM} y_{\rm diag}^{(d)} V_{\rm R}^{(d)}\right)_{ij}
\phi d_{{\rm R}j} + {\rm h.c.} 
\nonumber \\
\hspace{-1.2cm}
&~& ~~~~~~~~~~~~~~~\! 
= - y_{ij}^{(u)} \overline{q}_{{\rm L}i} \tilde{\phi} u_{{\rm R}j}
- y_{ij}^{(d)} \overline{q}_{{\rm L}i} \phi d_{{\rm R}j} + {\rm h.c.},
\label{Lq-Y}
\end{eqnarray}
and then Yukawa interactions in ${\mathscr{L}}_{\rm SM}^{\rm quark}$ are obtained.

In this way, we can set a goal to obtain the quark kinetic coefficients 
given in eqs.~\eqref{kq} -- \eqref{kd} or its equivalent ones,
under the assumption that ${\mathscr{L}'}_{\rm SM}^{\rm quark}$ 
effectively describes a relic from emergent kinetic terms
as a physics beyond the SM.
Then, we need kinetic coefficients with huge values,
because the eigenvalues of $\left(y_{\rm diag}^{(u)-1}\right)^2$ 
and $\left(y_{\rm diag}^{(d)-1}\right)^2$ are roughly estimated
at the weak scale as~\cite{PDG} 
\begin{eqnarray}
&~& \left(y_{\rm diag}^{(u)-1}\right)^2 \fallingdotseq
{\rm diag}\left(6.9 \times 10^{9},~ 1.9 \times 10^{4},~ 1.0\right),
\label{yu-diag-2-value}\\
&~& \left(y_{\rm diag}^{(d)-1}\right)^2 \fallingdotseq
{\rm diag}\left(1.4 \times 10^{9},~ 
3.4 \times 10^{6},~ 1.7 \times 10^{3}\right).
\label{yd-diag-2-value}
\end{eqnarray}
We look into how the flavor structure can be induced in the next section.

\section{`Emergent' kinetic terms}

To produce the quark mass hierarchy,
it is needed that kinetic coefficients can possess 
a hierarchy with huge values when they are diagonalized.
It can be realized based on the emergence proposal
that fermion kinetic terms in the SM can be generated radiatively
by loop corrections involving towers of massive states~\cite{CHI,CHI2}.

First, we give some basic assumptions.
(a) The SM fermions have no kinetic terms in the UV region.
(b) Yukawa interactions among the SM fields exist, and the SM fermion fields
are defined by making Yukawa interactions the canonical types.
(c) Towers of massive states exist with canonical kinetic terms.
(d) The SM fermions strongly couple to towers of massive states.
(e) The SM fermion kinetic terms (including gauge bosons 
via gauge interactions) appear emergently.

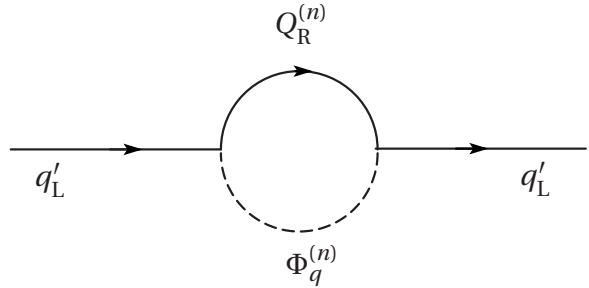
\begin{figure}[hbtp]
\begin{center}
\unitlength 0.1in
\begin{picture}( 30.1000, 12.7000)(  5.0000,-16.1000)
%
\special{pn 13}%
\special{pa 500 1060}%
\special{pa 1600 1060}%
\special{fp}%
%
\special{pn 13}%
\special{ar 2010 1060 410 410  3.1415927 6.2831853}%
%
\special{pn 13}%
\special{ar 2010 1080 410 410  6.2831853 6.4295268}%
\special{ar 2010 1080 410 410  6.5173316 6.6636731}%
\special{ar 2010 1080 410 410  6.7514780 6.8978195}%
\special{ar 2010 1080 410 410  6.9856243 7.1319658}%
\special{ar 2010 1080 410 410  7.2197707 7.3661121}%
\special{ar 2010 1080 410 410  7.4539170 7.6002585}%
\special{ar 2010 1080 410 410  7.6880634 7.8344048}%
\special{ar 2010 1080 410 410  7.9222097 8.0685512}%
\special{ar 2010 1080 410 410  8.1563560 8.3026975}%
\special{ar 2010 1080 410 410  8.3905024 8.5368438}%
\special{ar 2010 1080 410 410  8.6246487 8.7709902}%
\special{ar 2010 1080 410 410  8.8587951 9.0051365}%
\special{ar 2010 1080 410 410  9.0929414 9.2392829}%
\special{ar 2010 1080 410 410  9.3270877 9.4247780}%
%
\special{pn 13}%
\special{pa 2410 1056}%
\special{pa 3510 1056}%
\special{fp}%
\put(6.2000,-13.1000){\makebox(0,0)[lb]{$q'_{\rm L}$}}%
\put(31.6000,-13.0500){\makebox(0,0)[lb]{$q'_{\rm L}$}}%
\put(18.8000,-5.1000){\makebox(0,0)[lb]{$Q^{(n)}_{{\rm R}}$}}%
\put(19.4000,-17.8000){\makebox(0,0)[lb]{$\Phi^{(n)}_q$}}%
%
\special{pn 13}%
\special{pa 1020 1060}%
\special{pa 1170 1060}%
\special{fp}%
\special{sh 1}%
\special{pa 1170 1060}%
\special{pa 1104 1040}%
\special{pa 1118 1060}%
\special{pa 1104 1080}%
\special{pa 1170 1060}%
\special{fp}%
%
\special{pn 13}%
\special{pa 2830 1056}%
\special{pa 2980 1056}%
\special{fp}%
\special{sh 1}%
\special{pa 2980 1056}%
\special{pa 2914 1036}%
\special{pa 2928 1056}%
\special{pa 2914 1076}%
\special{pa 2980 1056}%
\special{fp}%
%
\special{pn 13}%
\special{pa 1990 650}%
\special{pa 2060 650}%
\special{fp}%
\special{sh 1}%
\special{pa 2060 650}%
\special{pa 1994 630}%
\special{pa 2008 650}%
\special{pa 1994 670}%
\special{pa 2060 650}%
\special{fp}%
\end{picture}%
\end{center}
\caption[1loop]{One-loop diagram to induce the kinetic term of $q'_{\rm L}$.}
\label{1loop}
\end{figure}

Let us study a simple model to grasp a feature of our proposal
and see if a realistic flavor structure can be generated or not.
The model has Yukawa interactions among each $q'_{\rm L}$, $u'_{\rm R}$ and $d'_{\rm R}$
and massive particles such that
\begin{eqnarray}
f^{(n)}_{Q~\!ij}\overline{Q}_{{\rm R}i}^{(n)} {\Phi}_q^{(n)} {q}'_{{\rm L}j},~~
f^{(n)}_{U~\!ij}\overline{U}_{{\rm L}i}^{(n)} {\Phi}_u^{(n)} u'_{{\rm R}j},~~
f^{(n)}_{D~\!ij}\overline{D}_{{\rm L}i}^{(n)} {\Phi}_d^{(n)} d'_{{\rm R}j},
\label{Y-KK}
\end{eqnarray}
where $f^{(n)}_{Q~\!ij}$, $f^{(n)}_{U~\!ij}$ and $f^{(n)}_{D~\!ij}$ are 
Yukawa coupling matrices, $i, j=1,2,3$,
${Q}_{{\rm R}i}^{(n)}$, ${U}_{{\rm L}i}^{(n)}$ and ${D}_{{\rm L}i}^{(n)}$
are massive fermions and
${\Phi}_q^{(n)}$, ${\Phi}_u^{(n)}$ and ${\Phi}_d^{(n)}$ are massive scalar fields.
We take quantized couplings $f^{(n)}_{Q~\!ij}$, $f^{(n)}_{U~\!ij}$ and $f^{(n)}_{D~\!ij}$ 
and quantized masses of
${Q}_{{\rm R}i}^{(n)}$, ${U}_{{\rm L}i}^{(n)}$ and ${D}_{{\rm L}i}^{(n)}$
such that
\begin{eqnarray}
&~& f^{(n)}_{Q~\!ij} = n_{Qi} f_{Q~\!ij},~~
f^{(n)}_{U~\!ij} = n_{Ui} f_{U~\!ij},~~
f^{(n)}_{U~\!ij} = n_{Di} f_{D~\!ij},
\label{f(n)}\\
&~& m^{(n)}_{Qi} = n_{Qi}m_{Qi},~~
m^{(n)}_{Ui} = n_{Ui}m_{Ui},~~
m^{(n)}_{Di} = n_{Di}m_{Di},
\label{m(n)}
\end{eqnarray}
where $n_{Qi}$, $n_{Ui}$ and $n_{Di}$ are integers
and no summation is done for repeated indices.\footnote{
We assume that interactions given by eq.~\eqref{Y-KK} also have a feature 
relating to universality which gauge interactions do, i.e.,
Kaluza-Klein modes can have gauge quantum numbers proportional to
their masses upon $S^1$ compactification.
This feature can be realized 
if interactions originated from a large unified gauge interaction.}
Here and hereafter, we treat 
$f_{Q~\!ij}$, $f_{U~\!ij}$ and $f_{D~\!ij}$ as complex matrices
with elements of $O(1)$, although one of them can become a diagonal form
after performing a suitable bi-unitary transformation, 
keeping both quark Yukawa interactions in the SM and kinetic terms of massive particles
the canonical ones.
Then, the kinetic coefficients of $q'_{{\rm L}}$ at the one-loop level
are calculated based on the diagram in Figure~\ref{1loop}.
After summing over towers of states up to the UV cutoff scale $\Lambda$, 
we estimate the contribution of $k_{ij}^{(q)}$ as
\begin{eqnarray}
&~& k_{ij}^{(q)} \simeq \sum_{n_{Qk}=1}^{N_{Qk}} \sum_{k=1}^3
f_{Q~\!ik}^{\dagger}~n_{Qk}^2 \ln\frac{\Lambda^2}{m^{(n)2}_{Qk}}~f_{Q~\!kj}
= \sum_{k=1}^3 f_{Q~\!ik}^{\dagger}~
\left(\sum_{n_{Qk}=1}^{N_{Qk}}n_{Qk}^2 \ln\frac{\Lambda^2}{m^{(n)2}_{Qk}}\right)
~f_{q~\!kj}
\nonumber \\
&~& ~~~~~~~~\! \simeq \sum_{k=1}^3 f_{Q~\!ik}^{\dagger}~N_{Qk}^3~f_{Q~\!kj}
\simeq \sum_{k=1}^3 f_{Q~\!ik}^{\dagger}~
\left(\frac{\Lambda}{m_{Qk}}\right)^{3}~f_{Q~\!kj},
\label{kq-est}
\end{eqnarray}
where we use the relation $\Lambda \simeq N_{Qk}m_{Qk}$ ($k=1,2,3$, no summation over $k$)
and $A \simeq B$ means $a = O(b)$ ($a$ and $b$ are values of $A$ and $B$, respectively).
Here, we assume that the one-loop contributions are dominant 
to generate the SM fermion kinetic terms in the IR region, 
and renormalized Kaluza-Klein propagators and coupling constants are used.
It is based on the fact that higher loop diagrams connected by the SM fermion propagators 
do not contribute in the absent of the SM fermion kinetic terms and 
a conjecture that the dynamics in the UV region
can be well-controlled by a topological nature of a fundamental theory, 
even in a strong coupling regime.

In the same way, we obtain the following type of kinetic coefficients:
\begin{eqnarray}
k_{ij}^{(q)} = \left(f_Q^{\dagger} \xi_q^2 f_Q\right)_{ij},~~
k_{ij}^{(u)} = \left(f_U^{\dagger} \xi_u^2 f_U\right)_{ij},~~
k_{ij}^{(d)} = \left(f_D^{\dagger} \xi_d^2 f_D\right)_{ij},
\label{kfxi}
\end{eqnarray}
where $\xi_q$, $\xi_u$ and $\xi_d$ are positive diagonal matrices given by
\begin{eqnarray}
&~& \xi_q \simeq {\rm diag}\left(\left(\frac{\Lambda}{m_{Q1}}\right)^{3/2},~~
\left(\frac{\Lambda}{m_{Q2}}\right)^{3/2},~~
\left(\frac{\Lambda}{m_{Q3}}\right)^{3/2}\right),
\label{xiq}\\
&~& \xi_u \simeq {\rm diag}\left(\left(\frac{\Lambda}{m_{U1}}\right)^{3/2},~~
\left(\frac{\Lambda}{m_{U2}}\right)^{3/2},~~
\left(\frac{\Lambda}{m_{U3}}\right)^{3/2}\right),
\label{xiu}\\
&~& \xi_d \simeq {\rm diag}\left(\left(\frac{\Lambda}{m_{D1}}\right)^{3/2},~~
\left(\frac{\Lambda}{m_{D2}}\right)^{3/2},~~
\left(\frac{\Lambda}{m_{D3}}\right)^{3/2}\right),
\label{xid}
\end{eqnarray}
respectively.
Then, the kinetic coefficients are rewritten as
\begin{eqnarray}
k^{(q)}_{ij} = \left({X'}_{q}^{\dagger}X'_{q}\right)_{ij},~~
k^{(u)}_{ij} = \left({X'}_{u}^{\dagger}X'_{u}\right)_{ij},~~
k^{(d)}_{ij} = \left({X'}_{d}^{\dagger}X'_{d}\right)_{ij},
\label{kN'}
\end{eqnarray}
using $3 \times 3$ complex matrices $X'_q$, $X'_u$ and $X'_d$ parametrized by
\begin{eqnarray}
X'_q = {V_q}^{\dagger} \xi_q f_Q,~~ 
X'_u = {V_u}^{\dagger} \xi_u f_U,~~
X'_d = {V_d}^{\dagger} \xi_d f_D,
\label{N'}
\end{eqnarray}
respectively.
Here, $V_q$, $V_u$ and $V_d$ are $3 \times 3$ unitary matrices.
In terms of $q_{\rm L} = X'_q q'_{\rm L}$, $u_{\rm R} = X'_u u'_{\rm R}$
and $d_{\rm R} = X'_d d'_{\rm R}$,
the canonical Yukawa interactions are rewritten as
\begin{eqnarray}
\hspace{-1.2cm}
&~& \mathscr{L}_{\rm CY}^{\rm quark} 
=  - \overline{q}'_{{\rm L}i} \tilde{\phi} u'_{{\rm R}i}
- \overline{q}'_{{\rm L}i} \phi d'_{{\rm R}i} + {\rm h.c.} 
\nonumber \\
\hspace{-1.2cm}
&~& ~~~~~~~~~~~~~~~\! 
= - \overline{q}_{{\rm L}i} \left(\left({X'}_q^{\dagger}\right)^{-1} {X'}_u^{-1}\right)_{ij}
\tilde{\phi}u_{{\rm R}j}
- \overline{q}_{{\rm L}i} \left(\left({X'}_q^{\dagger}\right)^{-1} {X'}_d^{-1}\right)_{ij}
\phi d_{{\rm R}j} + {\rm h.c.} 
\nonumber \\
\hspace{-1.2cm}
&~& ~~~~~~~~~~~~~~~\! 
= - \overline{q}_{{\rm L}i} 
\left({V_q}^{\dagger} \xi_q^{-1} (f_Q^{-1})^{\dagger} f_U^{-1} \xi_u^{-1} V_u\right)_{ij}
\tilde{\phi}u_{{\rm R}j}
- \overline{q}_{{\rm L}i} 
\left({V_q}^{\dagger} \xi_q^{-1} (f_Q^{-1})^{\dagger} f_D^{-1} \xi_d^{-1} V_d\right)_{ij}
\phi d_{{\rm R}j} 
\nonumber \\
\hspace{-1.2cm}
&~& ~~~~~~~~~~~~~~~~~~~
+ {\rm h.c.}.
\label{Lq-Yf}
\end{eqnarray}

Now, the kinetic terms take a canonical form with the global 
${\rm U}(3) \times {\rm U}(3) \times {\rm U}(3)/{\rm U}(1)$ symmetry,
and hence we have Yukawa coupling matrices such as
\begin{eqnarray}
y^{(u)}_{ij} = \left(\xi_q^{-1} (f_Q^{-1})^{\dagger} f_U^{-1} \xi_u^{-1}\right)_{ij},~~
y^{(d)}_{ij} = \left(\xi_q^{-1} (f_Q^{-1})^{\dagger} f_D^{-1} \xi_d^{-1}\right)_{ij},
\label{yxf}
\end{eqnarray}
by replacing $V_q q_{\rm L}$, $V_u u_{\rm R}$ and $V_d d_{\rm R}$
with $q_{\rm L}$, $u_{\rm R}$ and $d_{\rm R}$, respectively.

Let us consider the case that complex matrices $(f_Q^{-1})^{\dagger} f_U^{-1}$ and 
$(f_Q^{-1})^{\dagger} f_D^{-1}$ have non-vanishing elements of $O(1)$
and assume that there exist hierarchies such that 
$\xi_{q1}^{-1} \ll \xi_{q2}^{-1} \ll \xi_{q3}^{-1}$ and
$\xi_{u1}^{-1} \ll \xi_{u2}^{-1} \ll \xi_{u3}^{-1}$
because of a large mass difference in the up-type quark sector
and a small flavor mixing.
Then, $y^{(u)}_{ij}$ and $y^{(d)}_{ij}$ are approximated by the formulas:\footnote{
The structure of Yukawa coupling matrices resembles that derived from
the Froggatt-Nielsen mechanism~\cite{FN}.}
\begin{eqnarray}
y^{(u)}_{ij} \simeq \xi_{qi}^{-1} \xi_{uj}^{-1},~~
y^{(d)}_{ij} \simeq \xi_{qi}^{-1} \xi_{dj}^{-1}
\label{yxij}
\end{eqnarray}
and, after performing suitable bi-unitary transformations,
they can be diagonalized as
\begin{eqnarray}
&~& y^{(u)}_{\rm diag} = {V_{\rm L}^{(u)}}^{\dagger}y^{(u)}V_{\rm R}^{(u)} 
\simeq {\rm diag}\left(\xi_{q1}^{-1} \xi_{u1}^{-1},~,
\xi_{q2}^{-1} \xi_{u2}^{-1},~\xi_{q3}^{-1} \xi_{u3}^{-1}\right),~~
\label{yux-diag}\\
&~& y^{(d)}_{\rm diag} = {V_{\rm L}^{(d)}}^{\dagger}y^{(d)}V_{\rm R}^{(d)}
 \simeq {\rm diag}\left(\xi_{q1}^{-1} \tilde{\xi}_{d1}^{-1},~,
\xi_{q2}^{-1} \tilde{\xi}_{d2}^{-1},~\xi_{q3}^{-1} \tilde{\xi}_{d3}^{-1}\right),
\label{ydx-diag}
\end{eqnarray}
where $\tilde{\xi}_{d1}^{-1}$, $\tilde{\xi}_{d2}^{-1}$ and $\tilde{\xi}_{d3}^{-1}$
are some positive numbers.
Note that $\tilde{\xi}_{di}^{-1}$ $(i=1,2.3)$ do not necessarily 
agree with ${\xi}_{di}^{-1}$
because a hierarchy such as 
$\xi_{d1}^{-1} \ll \xi_{d2}^{-1} \ll \xi_{d3}^{-1}$ is not assumed.
The unitary matrices are given by
\begin{eqnarray}
&~& V_{\rm L}^{(u)},~~ V_{\rm L}^{(d)} \simeq 
\left(
\begin{array}{ccc}
1 & \xi_{q1}^{-1}/\xi_{q2}^{-1} & \xi_{q1}^{-1}/\xi_{q3}^{-1} \\
\xi_{q1}^{-1}/\xi_{q2}^{-1} & 1 & \xi_{q2}^{-1}/\xi_{q3}^{-1} \\
\xi_{q1}^{-1}/\xi_{q3}^{-1} & \xi_{q2}^{-1}/\xi_{q3}^{-1} & 1
\end{array} 
\right),~~
\label{VL}\\
&~& V_{\rm R}^{(u)} \simeq 
\left(
\begin{array}{ccc}
1 & \xi_{u1}^{-1}/\xi_{u2}^{-1} & \xi_{u1}^{-1}/\xi_{u3}^{-1} \\
\xi_{u1}^{-1}/\xi_{u2}^{-1} & 1 & \xi_{u2}^{-1}/\xi_{u3}^{-1} \\
\xi_{u1}^{-1}/\xi_{u3}^{-1} & \xi_{u2}^{-1}/\xi_{u3}^{-1} & 1
\end{array} 
\right).
\label{VRu}
\end{eqnarray}
The expression of $V_{\rm R}^{(d)}$ is not determined without specifying 
a magnitude relationship among ${\xi}_{d1}^{-1}$, ${\xi}_{d2}^{-1}$ and ${\xi}_{d3}^{-1}$
or giving explicit values, either.
From $y^{(u)}_{\rm diag} = {\rm diag}(y_u, y_c, y_t)$,
$y^{(d)}_{\rm diag} = {\rm diag}(y_d, y_s, y_b)$ and eqs.~\eqref{yux-diag}
and \eqref{ydx-diag}, we obtain the relations:
\begin{eqnarray}
y_u \simeq \xi_{q1}^{-1} \xi_{u1}^{-1},~~
y_c \simeq \xi_{q2}^{-1} \xi_{u2}^{-1},~~ y_t \simeq \xi_{q3}^{-1} \xi_{u3}^{-1},~~
y_d \simeq \xi_{q1}^{-1} \tilde{\xi}_{d1}^{-1},~~
y_s \simeq \xi_{q2}^{-1} \tilde{\xi}_{d2}^{-1},~~ y_b \simeq \xi_{q3}^{-1} \tilde{\xi}_{d3}^{-1}.
\label{xi-y}
\end{eqnarray}
From eqs.~\eqref{VKM} and \eqref{VL}, we obtain the relations:
\begin{eqnarray}
&~& (V_{\rm CKM})_{11} = V_{ud} \simeq 1,~~ (V_{\rm CKM})_{22} =V_{cs} \simeq 1,~~
(V_{\rm CKM})_{33} = V_{tb} \simeq 1,~~
\label{xi-VKMii}\\
&~& (V_{\rm CKM})_{12} = V_{us} \simeq \frac{\xi_{q1}^{-1}}{\xi_{q2}^{-1}},~~
(V_{\rm CKM})_{13} = V_{ub} \simeq \frac{\xi_{q1}^{-1}}{\xi_{q3}^{-1}},~~
(V_{\rm CKM})_{23} = V_{cb} \simeq \frac{\xi_{q2}^{-1}}{\xi_{q3}^{-1}},
\label{xi-VKMij}\\
&~& (V_{\rm CKM})_{21} = V_{cd} \simeq \frac{\xi_{q1}^{-1}}{\xi_{q2}^{-1}},~~
(V_{\rm CKM})_{31} = V_{td} \simeq \frac{\xi_{q1}^{-1}}{\xi_{q3}^{-1}},~~
(V_{\rm CKM})_{32} = V_{ts} \simeq \frac{\xi_{q2}^{-1}}{\xi_{q3}^{-1}}
\label{xi-VKMji}
\end{eqnarray}
and derive the relations $(V_{\rm CKM})_{ij} \simeq (V_{\rm CKM})_{ji}$
and $(V_{\rm CKM})_{13} \simeq (V_{\rm CKM})_{12}(V_{\rm CKM})_{23}$.

From eqs.~\eqref{xi-y} and \eqref{xi-VKMij}, the $\xi_{q1}^{-1}$, $\xi_{q2}^{-1}$,
$\xi_{u1}^{-1}$, $\cdots$, $\tilde{\xi}_{d2}^{-1}$ and $\tilde{\xi}_{d3}^{-1}$ are expressed as
\begin{eqnarray}
&~& \xi_{q1}^{-1} \simeq (V_{\rm CKM})_{13} \xi_{q3}^{-1},~~
\xi_{q2}^{-1} \simeq (V_{\rm CKM})_{23} \xi_{q3}^{-1},~~ 
\label{xi-q}\\
&~& \xi_{u1}^{-1} \simeq \frac{y_u}{(V_{\rm CKM})_{13}} \xi_{q3},~~
\xi_{u2}^{-1} \simeq \frac{y_c}{(V_{\rm CKM})_{23}} \xi_{q3},~~
\xi_{u3}^{-1} \simeq y_t \xi_{q3},~~
\label{xi-u}\\
&~& \tilde{\xi}_{d1}^{-1} \simeq \frac{y_d}{(V_{\rm CKM})_{13}} \xi_{q3},~~
\tilde{\xi}_{d2}^{-1} \simeq \frac{y_s}{(V_{\rm CKM})_{23}} \xi_{q3},~~
\tilde{\xi}_{d3}^{-1} \simeq y_b \xi_{q3},
\label{xi-d}
\end{eqnarray}
using observable parameters and $\xi_{q3}$.
Because $y_t = O(1)$, $\xi_{qi}^{-1} \le O(1)$ and $\xi_{ui}^{-1} \le O(1)$ hold,
the magnitude of $\xi_{q3}^{-1}$ and $\xi_{u3}^{-1}$ 
should be of $O(1)$ from $y_t \simeq \xi_{q3}^{-1} \xi_{u3}^{-1}$.
Furthermore, from eqs.~\eqref{xiq}, \eqref{xiu},
\eqref{xi-q} and \eqref{xi-u}, we obtain the relations:
\begin{eqnarray}
&~& m_{Q1} \simeq (V_{\rm CKM})_{13}^{2/3} \times \Lambda,~~
m_{Q2} \simeq (V_{\rm CKM})_{23}^{2/3} \times \Lambda,~~
m_{Q3} \simeq \Lambda,~~
\label{mQ}\\
&~& m_{U1} \simeq \left(\frac{y_u}{(V_{\rm CKM})_{13}}\right)^{2/3} \times \Lambda,~~
m_{U2} \simeq \left(\frac{y_c}{(V_{\rm CKM})_{23}}\right)^{2/3} \times \Lambda,~~
m_{U3} \simeq \Lambda.
\label{mU}
\end{eqnarray}

For reference, using the values of observable parameters at the weak scale~\cite{PDG}:
\begin{eqnarray}
&~& y_u \fallingdotseq 1.2 \times 10^{-5},~~
y_c \fallingdotseq 7.3 \times 10^{-3},~~
y_t \fallingdotseq 1.0,~~
\label{yu-diag-value}\\
&~& y_d \fallingdotseq 2.7 \times 10^{-5},~~
y_s \fallingdotseq 5.4 \times 10^{-4},~~
y_b \fallingdotseq 2.4 \times 10^{-2},~~
\label{yd-diag-value}\\
&~& |(V_{\rm CKM})_{ij}| \fallingdotseq
\left(
\begin{array}{ccc}
0.97435 & 0.225 & 0.00369 \\
0.22486 & 0.97349 & 0.04182 \\
0.00857 & 0.0411 & 0.999118
\end{array} 
\right),
\label{VKM-value}
\end{eqnarray}
we estimate the magnitude of $\xi_{q1}^{-1}$, $\xi_{q2}^{-1}$,
$\xi_{u1}^{-1}$, $\cdots$, $\tilde{\xi}_{d2}^{-1}$ and $\tilde{\xi}_{d3}^{-1}$ as
\begin{eqnarray}
&~& \xi_{q1}^{-1} \simeq 3.7~(8.6) \times 10^{-3},~~
\xi_{q2}^{-1} \simeq 4.2~(4.1) \times 10^{-2},~~ 
\xi_{q3}^{-1} \simeq 1.0,
\label{xi-q-est}\\
&~& \xi_{u1}^{-1} \simeq 3.3~(1.4) \times 10^{-3},~~
\xi_{u2}^{-1} \simeq 1.7~(1.8) \times 10^{-1},~~
\xi_{u3}^{-1} \simeq 1.0,~~
\label{xi-u-est}\\
&~& \tilde{\xi}_{d1}^{-1} \simeq 7.3~(3.1) \times 10^{-3},~~
\tilde{\xi}_{d2}^{-1} \simeq 1.3~(1.3) \times 10^{-2},~~
\tilde{\xi}_{d3}^{-1} \simeq 2.4 \times 10^{-2},
\label{xi-d-est}
\end{eqnarray}
where the numbers in the parentheses are obtained by using $(V_{\rm CKM})_{31}$
and $(V_{\rm CKM})_{32}$ in place of $(V_{\rm CKM})_{13}$ and $(V_{\rm CKM})_{23}$, respectively. 
From eqs.~\eqref{xi-q-est} and \eqref{xi-u-est},
we find that the assumption $\xi_{q1}^{-1} \ll \xi_{q2}^{-1} \ll \xi_{q3}^{-1}$ and
$\xi_{u1}^{-1} \ll \xi_{u2}^{-1} \ll \xi_{u3}^{-1}$ hold to some extent.
Furthermore, we estimate the magnitude of $m_{Q1}$, $\cdots$, $m_{D2}$ and $m_{D3}$ as
\begin{eqnarray}
\hspace{-1cm}
&~& m_{Q1} \simeq 2.4~(4.2) \times 10^{-2} \times \Lambda,~~
m_{Q2} \simeq 1.2~(1.2) \times 10^{-1} \times \Lambda,~~
m_{Q3} \simeq \Lambda,~~
\label{mQ-value}\\
\hspace{-1cm}
&~& m_{U1} \simeq 2.2~(1.3) \times 10^{-2} \times \Lambda,~~
m_{U2} \simeq 3.1~(3.2) \times 10^{-1} \times \Lambda,~~
m_{U3} \simeq \Lambda,~~
\label{mU-value}\\
\hspace{-1cm}
&~& m_{D1} \simeq 3.8~(2.2) \times 10^{-2} \times \Lambda,~~
m_{D2} \simeq 5.5~(5.5) \times 10^{-2} \times \Lambda,~~
m_{D3} \simeq 8.3 \times 10^{-2} \Lambda,
\label{mD-value}
\end{eqnarray}
where the numbers in the parentheses are obtained by using $(V_{\rm CKM})_{31}$
and $(V_{\rm CKM})_{32}$ in place of $(V_{\rm CKM})_{13}$ and $(V_{\rm CKM})_{23}$, respectively,
and we take $\tilde{\xi}_{di}^{-1} = \xi_{di}^{-1}$ for our guidance.
Note that the relations $(V_{\rm CKM})_{ij} \simeq (V_{\rm CKM})_{ji}$
and $(V_{\rm CKM})_{13} \simeq (V_{\rm CKM})_{12}(V_{\rm CKM})_{23}$ are consistent with
the experimental data \eqref{VKM-value}.   
Physical parameters, in general, receive radiative corrections, 
and the above values should be evaluated by considering renormalization effects.
In any case, we understand that the quark mass hierarchy and flavor mixing can originate from
a milder mass hierarchy on massive fermions. 

Finally, we point out that kinetic coefficients can become specific forms
including democratic-type matrices by a change of variables.
After changing variables as
\begin{eqnarray}
q'^{({\rm de})}_{\rm L} = U^{({\rm de})} f_Q q'_{\rm L},~~
u'^{({\rm de})}_{\rm R} = U^{({\rm de})} f_U u'_{\rm R},~~
d'^{({\rm de})}_{\rm R} = U^{({\rm de})} f_D d'_{\rm R},
\label{q'de}
\end{eqnarray}
the kinetic coefficients $k^{(q)}$, $k^{(u)}$ and $k^{(d)}$ become
\begin{eqnarray}
&~& k^{(q)} = \xi_{q1}^2 S^{(1)} + \xi_{q2}^2 S^{(2)} + \xi_{q3}^2 S^{(3)},~~
\label{kq-D}\\
&~& k^{(u)} = \xi_{u1}^2 S^{(1)} + \xi_{u2}^2 S^{(2)} + \xi_{u3}^2 S^{(3)},~~
\label{ku-D}\\
&~& k^{(d)} = \xi_{d1}^2 S^{(1)} + \xi_{d2}^2 S^{(2)} + \xi_{d3}^2 S^{(3)},
\label{kd-D}
\end{eqnarray}
where $\xi_{qi}^2$, $\xi_{ui}^2$ and $\xi_{di}^2$ are 
the $(i~i)$ elements in $\xi_{q}^2$, $\xi_{u}^2$ and $\xi_{d}^2$, respectively.
The $U^{({\rm de})}$, $S^{(1)}$, $S^{(2)}$ and $S^{(3)}$ are $3 \times 3$ matrices given by
\begin{eqnarray}
\hspace{-1.2cm}
&~& U^{({\rm de})} = \frac{1}{\sqrt{3}}
\left(
\begin{array}{ccc}
1 & 1 & 1 \\
1 & \omega & \omega^2 \\
1 & \omega^2 & \omega
\end{array} 
\right),~~
\label{UD}\\
\hspace{-1.2cm}
&~& 
S^{(1)} = \frac{1}{3}
\left(
\begin{array}{ccc}
1 & 1 & 1 \\
1 & 1 & 1 \\
1 & 1 & 1
\end{array} 
\right),~~
S^{(2)} = \frac{1}{3}
\left(
\begin{array}{ccc}
1 & \omega^2 & \omega \\
\omega & 1 & \omega^2 \\
\omega^2 & \omega & 1
\end{array} 
\right),~~
S^{(3)} = \frac{1}{3}
\left(
\begin{array}{ccc}
1 & \omega & \omega^2 \\
\omega^2 & 1 & \omega \\
\omega & \omega^2 & 1
\end{array} 
\right),
\label{S}
\end{eqnarray}
where $\omega = e^{\frac{2}{3}\pi i}$ and $S^{(1)}$ is a democratic-type matrix.

\section{Conclusions and discussions}

We have studied the flavor structure of quarks in the SM
from a viewpoint of a canonical type of Yukawa interactions 
and an emergence of kinetic terms.
We have found that a realistic structure can be generated based on the emergence proposal 
that quark kinetic terms appear in the IR region,
as a result of radiative corrections
involving towers of massive states,
and the quark mass hierarchy and flavor mixing can originate from
a milder mass hierarchy on massive fermions.
A similar analysis can be applied 
to find out the origin of the lepton flavor structure~\cite{CHI2}.

There are several problems that remains to be solved.
Why do the SM fields have no kinetic terms in the UV region?
It might be caused by some topological nature of a fundamental theory
including a quantum gravity~\cite{AGOV}.
In contrast, why do towers of massive fields have kinetic terms
in the UV region?
And what is the origin of such kinetic terms?
What is the origin of gauge fields and gauge interactions?
What is the origin of Yukawa interactions?
In the first place, what is the origin of quantum fields
including the SM particles and towers of massive states?
A fundamental theory such as superstring theory and/or M theory
is expected to answer the above questions.

\section*{Acknowledgments}
The author thanks N.~Haba and T.~Yamada for valuable discussions.
This work was supported in part by scientific grants from the Ministry of Education, Culture,
Sports, Science and Technology under Grant No.~22K03632.

\end{document}